\documentstyle{article}

\begin{document}
\setlength{\marginparwidth}{0cm}
\setlength{\marginparsep}{0cm}

\title{
{\LARGE \bf On the Connection between The Radial Momentum Operator and the Hamiltonian in \boldmath $ n$ Dimensions}}
\author{\large  Gil Paz $^{a)}$ \\ \normalsize\it Physics Department,
Technion-Israel Institute of Technology, 32000 Haifa, Israel}
\date{}
\maketitle
\begin{abstract}
\large The radial momentum operator in quantum mechanics is usually obtained through canonical quantization of the (symmetrical form of the) classical radial momentum. We show that the well known connection between the Hamiltonian of a free particle and the radial momentum operator $\hat{H}=\hat{P}_{r}^2/2m+
\mbox{\boldmath $ \hat{L}^2$}/2mr^{2}$ is true \bf only \rm  in 1 or 3 dimensions. In general, an extra term of the form $\hbar^{2}(n-1)(n-3)/ 2m \cdot 4r^{2}$ has to be added to the Hamiltonian.
\end{abstract}
\thispagestyle{empty}

\newpage
\setcounter{page}{1}
\large
\section{\large \bf INTRODUCTION}

In quantum mechanics the radial momentum operator in three dimensions 
arises through quantization of the symmetric form of the classical radial momentum$^{1}$:
\begin{equation}
      P _{r}=\frac{1}{2}(\mbox{\boldmath $ p\cdot \hat{r} + \hat{r} \cdot p$}).
\end{equation}

It has the form:
\begin{equation}
 \hat{P}_{r}=-i\hbar\left( \frac{\partial}{\partial r }+\frac{1}{r} \right)=
  -i\hbar\frac{1}{r}\frac{\partial}{\partial r }r
\end{equation}

and the desirable property that the Hamiltonian of a free particle can be written as:
\begin{equation}
 \hat{H}=\frac{-\hbar^2\nabla^2}{2m}=
\frac{\hat{P}_{r}^2}{2m}+\frac{\mbox{\boldmath $ \hat{L}^2$}}{2mr^2}.
\end{equation}
If we try to repeat the procedure in cylindrical coordinates we would get through quantization of the classical radial momentum:
\begin{equation}
P _{\rho}=\frac{1}{2}(\mbox{\boldmath $ p\cdot\hat{\rho}+\hat{\rho}\cdot p$}),
\end{equation}
the operator:
\begin{equation}
 \hat{P}_{\rho}=-i\hbar\left (\frac{\partial}{\partial \rho }+\frac{1}{2\rho}
\right).
\end{equation}
The Hamiltonian in cylindrical coordinates is:
\begin{equation}
 \hat{H}=\frac{-\hbar^2\nabla^2}{2m}=\frac{-\hbar^2}{2m} \left(
\frac{\partial^2}{\partial\rho^2}+\frac{1}{\rho}\frac{\partial}{\partial \rho}
\right)+\frac{\hat{L_{z}}}{2m\rho^2}+ \frac{\hat{P}_{z}^2}{2m}\mbox{\boldmath $\neq$} \frac{\hat{P}_{\rho}^2}{2m}+\frac{\hat{L_{z}}}{2m\rho^2}+ \frac{\hat{P}_{z}^2}{2m},
\end{equation}
Since
\begin{equation}
\frac{\hat{P}_{\rho}^2}{2m}=\frac{-\hbar^2}{2m} \left(\frac{\partial^2}{\partial\rho^2}+\frac{1}{\rho}\frac{\partial}{\partial \rho}-\frac{1}{4\rho^2}\right).
\end{equation}
We see that the Hamiltonian of a free particle in cylindrical coordinates can not be expressed in a ``simple'' form using the radial momentum. This fact is 
usually overlooked in Quantum Mechanics textbooks although it can be found
$^{2}$.

The aim of this paper is to explain the reason for the failure of 
the procedure in cylindrical coordinates. The crucial point is to notice that cylindrical coordinates are plane polar coordinates (that is polar coordinates in two dimensions) ``plus'' a Cartesian coordinate (``z'' coordinate). Since
quantization in Cartesian coordinates is a straight forward procedure, we are led to suspect that the problem that we have encountered lies in the ``polar''
nature of the cylindrical coordinates. 

The main difference between plane polar coordinates and spherical 
coordinates is the dimension, i.e. plane polar coordinates are polar coordinates in 2 dimensions while spherical coordinates are polar coordinates in 3 dimensions. Therefore it seems advisable to analyze the quantization 
procedure in $n$-dimensional polar coordinates.

In the next section we will see that the connection between the Hamiltonian of a free particle and the radial momentum operator $\hat{H}=\hat{P}_{r}^2/2m+\mbox{\boldmath $ \hat{L}^2$}/2mr^{2}$ is true \bf only \rm  in 1 or 3 dimensions. In general, an extra term of the form $\hbar^{2}(n-1)(n-3)/ 2m \cdot 4r^{2}$ has to be added to the Hamiltonian. Therefore the ``procedure'' failed due to the fact that cylindrical coordinates contain in them plane polar coordinates which are polar coordinates in two dimensions.

\newpage
\section{\large \bf \mbox{\boldmath$n$}-DIMENSIONAL CALCULATION}
In this section we will repeat the quantization procedure for polar coordinates in $n$-dimensional Euclidean space. First, we will derive the expression for the Hamiltonian of a free particle and then we will obtain the radial momentum operator through quantization of the symmetric form of the classical radial momentum in $n$ dimensions.

Whenever one is working with curvilinear coordinates like polar coordinates,
one should pay attention to the fact that the Jacobian of the transformation
from Cartesian coordinates could be zero. In polar coordinates the Jacobian is zero for $r=0$ in any number of dimensions. Therefore, in the following discussion it will be implicitly assumed that $r\neq 0$.

polar coordinates in $n$-dimensional Euclidean space are: $r,\theta_{n-2}\ldots,\theta_{1},\phi$. where $0\leq\phi\leq2\pi$ , $0\leq\theta_{i}\leq\pi$.
The relation between the polar coordinates and the Cartesian coordinates
$x_{1},\ldots,x_{n}$ is:
\begin{eqnarray}
x_{1}&=&r\cos\phi \sin\theta_{1}\sin\theta_{2}\cdots\sin\theta_{n-2} \nonumber \\
x_{2}&=&r\sin\phi \sin\theta_{1}\sin\theta_{2}\cdots\sin\theta_{n-2} \nonumber \\
x_{3}&=&r\cos \theta_{1}\sin\theta_{2}\cdots\sin\theta_{n-2}  \nonumber \\
\vdots \nonumber \\
x_{i}&=&r\cos\theta_{i-2}\sin\theta_{i-1}\cdots\sin\theta_{n-2} \nonumber\\
\vdots \nonumber \\
x_{n}&=&r\cos\theta_{n-2}. \nonumber\\
\end{eqnarray}
A length element is defined using the metric tensor:
\begin{equation}
ds^2=\sum_{i,j=1}^{n} g_{ij}du^{i}du^{j} ,
\end{equation}
where: $u^{1}=r$, $u^{2}=\theta_{n-2}$, \ldots , $u^{n-1}=\theta_{1}$, $u^{n}=\phi$. \\
The metric tensor in $n$-dimensional polar coordinates is:
\begin{equation}
g_{ij}=\left(\begin{array}{ccccc}
              1&&&& \\
              &r^{2}&&& \\
              &&r^{2}\sin^{2}\theta_{n-2}&& \\
              &&&\ddots& \\
              &&&&r^{2}\sin^{2}\theta_{n-2}\cdots\sin^{2}\theta_{1} \\
              \end{array}
                          \right)
.
\end{equation}
If we define: $g^{ij}=(g_{ij})^{-1}$ and
\begin{equation}
 g\equiv\mbox{\rm det } g_{ij} = r^{2(n-1)}\sin^{2(n-2)}\theta_{n-2}\cdots\sin^{2}\theta_{1},
\end{equation} 
we can write the Laplacian as$^{3}$:
\begin{equation}
\Delta=\frac{1}{\sqrt{g}}\sum_{i,j=1}^{n}\frac{\partial}{\partial u^{i} }
\left( \sqrt{g}g^{ij}\frac{\partial}{\partial u^{j}} \right). 
\end{equation}
Since we are interested in the part of the Laplacian which is a function of r alone , we need to look at the first term of the sum that is: $i=j=1$.\\
We get:
\begin{equation}
\left(\Delta\right)_{r}=\frac{1}{r^{n-1}}\frac{\partial}{\partial r} \left( r^{n-1} \frac{\partial}{\partial r} \right) = \frac{\partial^2}{\partial r^{2}}+
\frac{(n-1)}{r}\frac{\partial}{\partial r}.
\end{equation} 
Thus the Hamiltonian of a free particle of a mass $m$ is:
\begin{equation}
H=\frac{-\hbar^{2}}{2m}\Delta=\frac{-\hbar^{2}}{2m}\left[\frac{\partial^2}{\partial r^{2}}+
\frac{(n-1)}{r}\frac{\partial}{\partial r} +\mbox{``angular'' terms} \right]\; .
\end{equation}
 
The $n$-dimensional radial momentum operator is:

\begin{equation}
\hat{P}_{r} = \frac{-i\hbar}{2} \left[ \mbox{\boldmath $\nabla$}
\cdot(\mbox{\boldmath $\hat{r}$}\cdots)+\mbox{\boldmath $\hat{r}$}\cdot \mbox{\boldmath $\nabla$} \right] 
\end{equation}

Where \mbox{\boldmath $\nabla$} is $\left( \frac{\partial}{\partial x_{1}},\cdots,\frac{\partial}{\partial x_{n}} \right)$ and the notation 
$\mbox{\boldmath $\nabla$}\cdot(\mbox{\boldmath$\hat{r}$}\cdots)$
indicates that \mbox{\boldmath $\nabla$} differentiates \mbox{\boldmath$\hat{r}$} and everything on its right.\\
Using the identity:
\begin{equation}
\bf \mbox{\boldmath $\nabla$}(\mbox{\boldmath$\hat{r}$}\psi) = \mbox{\boldmath$\hat{r}$}\cdot \mbox{\boldmath $\nabla$}\psi+\psi \mbox{\boldmath $\nabla$}\cdot \mbox{\boldmath$\hat{r}$}
\end{equation}
we can write the operator as:
\begin{equation}
\hat{P}_{r} = \frac{-i \hbar}{2} \left[ 2 \bf \mbox{\boldmath$\hat{r}$}\cdot \mbox{\boldmath $\nabla$}+\mbox{\boldmath $\nabla$} \cdot(\mbox{\boldmath$\hat{r}$}) \right].
\end{equation}
In $n$-dimensions: 
\begin{equation}
{\bf \mbox{\boldmath$\hat{r}$}=}\frac{\mbox{\boldmath$r$}}{r}=\frac{\sum_{i=1}^{n}x_{i} {\mbox{\boldmath$\hat{x}_{i}$}}}
{(\sum_{i=1}^{n}x_{i}^{2})^{1/2}} \; ,
\end{equation}
and therefore:
\begin{equation}
\mbox{\boldmath $\nabla$}\cdot {\bf\mbox{\boldmath$\hat{r}$}}=\sum_{i=1}^{n}\frac{\partial}{\partial x_{i} } \left( \frac{x_{i}}{r}\right)=\sum_{i=1}^{n}[\frac{1}{r}+
x_{i}\left(\frac{-1}{r^{2}}\right) \frac{\partial r}{\partial x_{i}}]=
\frac{n}{r}-\frac{1}{r^{2}}\sum_{i=1}^{n}
x_{i}\frac{\partial r}{\partial x_{i}} \; .
\end{equation}
Since 
\begin{equation}
r^2=\sum_{i=1}^{n}x_{i}^{2} \Rightarrow 2r\frac{\partial r}{\partial x_{i}}=
2x_{i} \Rightarrow \frac{\partial r}{\partial x_{i}}=\frac{x_{i}}{r} \; ,
\end{equation}
we can write:
\begin{equation}
\mbox{\boldmath $\nabla$}\cdot {\bf\mbox{\boldmath$\hat{r}$}}=\frac{n}{r}-\frac{1}{r^{2}}\sum_{i=1}^{n}\frac{x_{i}^{2}}{r}=
\frac{n-1}{r} \; .
\end{equation}
By definition $\frac{\partial}{\partial r}=\bf \mbox{\boldmath$\hat{r}$}\cdot \mbox{\boldmath $\nabla$}$, so the radial momentum operator in $n$ dimensions is:
\begin{equation}
\hat{P}_{r} = -i\hbar \left( \frac{\partial}{\partial r}+\frac{n-1}{2r} \right)
\; .
\end{equation} 
Using the identity: 
\begin{equation}
\left [\frac{\partial}{\partial r},\frac{n-1}{2r}\right]=-\frac{(n-1)}{2r^2},
\end{equation}
we get:  
\begin{eqnarray}
\frac{\hat{P}_{r}^{2}}{2m}=-\frac{\hbar^2}{2m}\left\{ \frac{\partial^2}{\partial r^{2}} +2\cdot\frac{n-1}{2r}\frac{\partial}{\partial r}+ \left[ \frac{\partial}{\partial r},\frac{n-1}{2r} \right]+\frac{(n-1)^{2}}{4r^2} \right \}=\\ \nonumber
-\frac{\hbar^2}{2m}\left\{ \frac{\partial^2}{\partial r^{2}} +\frac{n-1}{r}\frac{\partial}{\partial r}+\frac{(n-1)(n-3)}{4r^{2}}
\right\} \; ,
\end{eqnarray}
which doesn't agree with (14) unless  $n = 1,3$.\\
It is possible to define $n$-dimensional angular momentum operators 
so that the Hamiltonian of a free particle in $n$ dimensions is$^{4}$:
\begin{equation}
H=\frac{-\hbar^{2}}{2m}\Delta=\frac{-\hbar^{2}}{2m}\left(\frac{\partial^2}{\partial r^{2}}+\frac{n-1}{r}\frac{\partial}{\partial r}\right)+\frac{\mbox{\boldmath $ \hat{L}^2$}}{ 2mr^2}\;,
\end{equation}
while
\begin{equation}
\frac{\hat{P}_{r}^{2}}{2m} + \frac{\mbox{\boldmath $ \hat{L}^2$}}{2mr^2}=
-\frac{\hbar^2}{2m}\left\{ \frac{\partial^2}{\partial r^{2}} +\frac{n-1}{r}\frac{\partial}{\partial r}+\frac{(n-1)(n-3)}{4r^{2}}
\right\}+\frac{\mbox{\boldmath $ \hat{L}^2$}}{2mr^2}\;,
\end{equation}
and the two expressions (25) and (26) are equal only if $n=1,3$.

\newpage
\section{\large \bf DISCUSSION AND CONCLUSIONS }
First we would like to comment that the form of the radial momentum operator, that we received through ``canonical'' quantization, is identical to the form which is widely accepted in the literature.$^{4,5,6}$

It might seem that a transformation of the form :
\begin{equation}
\hat{P}_{r}\rightarrow \hat{P}_{r}  -i\hbar f(r) \; ,
\end{equation}
could fix things. This kind of transformation does not change the commutation relations: $[r,\hat{P}_{r}]=i\hbar$ and reduces to the classical radial momentum in the classical limit. Under such transformation $\hat{P}_{r}$ turns into:
\begin{equation}
\hat{P}_{r}=-i\hbar\left( \frac{\partial}{\partial r}+\frac{n-1}{2r}+f(r) \right) \; , 
\end{equation}
but then:
\begin{equation}
\frac{\hat{P}_{r}^{2}}{2m}=
-\frac{\hbar^2}{2m}\left\{ \frac{\partial^2}{\partial r^{2}} +\left[\frac{n-1}{r}+2f(r) \right]\frac{\partial}{\partial r}+\left[\frac{(n-1)(n-3)}{4r^{2}}+(f(r))^{2}+\frac{df}{dr} \right]
\right\} \; .
\end{equation}
A comparison of the coefficients of $\frac{\partial}{\partial r}$ in (29) and (14) clearly shows that $f(r)$ has to be zero.

Therefore we come into the conclusion that in $n$-dimensional polar coordinates we have as the Hamiltonian of a free particle:
\begin{equation}
\hat{H}=\frac{\hat{P}_{r}^{2}}{2m} + \frac{\mbox{\boldmath $ \hat{L}^2$}}{2mr^2} 
+\frac{\hbar^2}{2m}\cdot \frac{(n-1)(n-3)}{4r^{2}}.
\end{equation}

Ess\'{e}n$^{7}$ has shown that when one quantize a classical Hamiltonian of the form $H(p,q)=p_{i}g^{ij}p_{j}/2$ the corresponding Hamiltonian operator is: $\hat{H}=\hat{p}_{i} g^{ij} \hat{p}_{j}/2+\Delta V_{D}$ where  $\hat{p}_{i} = 
-i \hbar g^{-1/4}  \frac{\partial}{\partial q^{i}}g^{1/4}$ and $\Delta V_{D}=-(\hbar^{2}/2)g^{-1/4}(g^{1/2}g^{ij}(g^{-1/4}),_{i}),_{j}$ .\\
In $n$-dimensional polar coordinates part of his correction is the term $\hbar^{2}(n-1)(n-3)/ 2m \cdot 4r^{2}$ that we mentioned earlier, and the rest is corrections to the momenta conjugate to the angular coordinates. But even when we
try to express the Hamiltonian as a function of the radial and angular momentum alone we face difficulties, as we have seen.  

It should be noted that  the fact  that polar coordinates in three dimensions has this special property has caused confusion before.$^{8}$

Finally it should be noted that the radial momentum operator has a very dubious status as an observable and it is not self adjoint.$^{9,10,11}$

In conclusion, we have seen that the well known connection between the Hamiltonian of a free particle and the radial momentum operator
\begin{displaymath}
\hat{H}=\frac{\hat{P}_{r}^2}{2m}+\frac{\mbox{\boldmath $ \hat{L}^2$}}{2mr^{2}}
\end{displaymath}
is true \bf only \rm  in 1 or 3 dimensions.\\
In general, the connection is :
\begin{displaymath}
\hat{H}=\frac{\hat{P}_{r}^{2}}{2m} + \frac{\mbox{\boldmath $ \hat{L}^2$}}{2mr^2} 
+\frac{\hbar^2}{2m}\cdot \frac{(n-1)(n-3)}{4r^{2}}
\end{displaymath}

Therefore, the analogy between the classical Hamiltonian  and the Quantum Hamiltonian for a free particle in spherical coordinates   is \bf purely \rm a ``numerical coincidence'' and not as fundamental as one is led to believe from many textbooks$^{6,12,13}$.

\newpage
\section*{\large \bf ACKNOWLEDGMENT}
This work was supported by the Technion Graduate School.\\\\
I thank Erez Berg for a fruitful discussion and Amnon Harel for his remarks. \\\\
a) E-mail address: gilpaz@tx.technion.ac.il\\ 
$^{1}$ R. L. Liboff, \it Introductory Quantum Mechanics \rm , $\rm 3^{rd}$ ed.  (Addison-Wesley, USA, 1998) p. 424-425\\
$^{2}$ R. L. Liboff, \it Introductory Quantum Mechanics \rm , $\rm 3^{rd}$ ed.  (Addison-Wesley, USA, 1998) p. 490-491\\
$^{3}$ M. R. Spiegel, \it Vector Analysis, \rm (McGraw-Hill, Great Britain, 1974) p. 174 \\
$^{4}$ J.D. Louck, J.Mol.Spec. \bf 4\rm, 298 (1960)\\
$^{5}$ B.S. deWitt, Phys.Rev. \bf 85\rm, 653 (1952)\\
$^{6}$ E. Merzbacher \it Quantum Mechanics \rm , $\rm 3^{rd}$ ed.
(John Wiley \& Sons, USA, 1998) p. 255\\
$^{7}$ H. Ess\'{e}n, Am.J.Phys. \bf 46\rm, 983 (1978)\\
$^{8}$ G.R. Gruber, Found. of Phys. \bf 1, \rm3, 227 (1971); G.R. Gruber, Inter.J.Theor.Phys. \bf 6, \rm1, 31 (1972) \\
$^{9}$ R.L. Liboff, I. Nebenzahl, H.H. Fleischmann,
 Am.J.Phys. \bf 41\rm, 976 (1973)\\  
$^{10}$ J.M. Domingos M.H. Caldeira, Found. of Phys. \bf 14, \rm 2, 147 (1984) \\
$^{11}$ G. Paz, math-ph/0009016 \\
$^{12}$ P.A.M Dirac,\it The Principles of Quantum Mechanics \rm , $\rm 4^{rd}$ ed. (Revised) (Oxford University Press, Hong Kong, 1995) p. 153\\
$^{13}$ A. Messiah, \it Quantum Mechanics \rm , (North-Holland, Amsterdam, 1965) p. 346-347\\

\end{document}